\documentclass[num-refs]{wiley-article}

\usepackage{siunitx}

\papertype{Original Article}
\paperfield{Journal Section}

\title{Understanding and adjusting for the selection bias from a proof-of-concept study to a more confirmatory study}

\abbrevs{ABC, a black cat; DEF, doesn't ever fret; GHI, goes home immediately.}

\author[1]{Yongming Qu PhD}
\author[1]{Yu Du PhD}
\author[1]{Ying Zhang PhD}
\author[1]{Lei Shen PhD}

\affil[1]{Department of Biometrics, Eli Lilly and Company, Indianapolis, Indiana, 46285, USA}

\corraddress{Yongming Qu PhD, Department of Biometrics, Eli Lilly and Company, Indianapolis, Indiana, 46285, USA}
\corremail{qu\_yongming@lilly.com}

\fundinginfo{No funding supports this research.}

\runningauthor{Qu, et al.}

\begin{document}

\maketitle

\begin{abstract}
It has long been noticed that the efficacy observed in small early phase studies is generally better than that observed in later larger studies. Historically, the inflation of the efficacy results from early proof-of-concept studies is either ignored, or adjusted empirically using a frequentist or Bayesian approach. In this article, we systematically explained the underlying reason for the inflation of efficacy results in small early phase studies from the perspectives of measurement error models and selection bias. A systematic method was built to adjust the early phase study results from both frequentist and Bayesian perspectives. A hierarchical model was proposed to estimate the distribution of the efficacy for a portfolio of compounds, which can serve as the prior distribution for the Bayesian approach. We showed through theory that the systematic adjustment provides an unbiased estimator for the true mean efficacy for a portfolio of compounds. The adjustment was applied to paired data for the efficacy in early small and later larger studies for a set of compounds in diabetes and immunology. After the adjustment, the bias in the early phase small studies seems to be diminished.
\end{abstract}

\section{Introduction}
\label{section_intro}
Clinical development of a new treatment is a lengthy, rigorous, and very costly process. Each treatment generally goes through a stage-wise procedure before regulatory approval into the market. Generally, research begins with smaller exploratory studies and gradually moves to larger confirmatory studies. 
It has been increasingly recognized that early phase trials can not only provide safety assessments, but also have the potential to evaluate the efficacy signal.
Sponsors tend to only select compounds that show promising efficacy (e.g. meeting a certain threshold value $\delta$) and a reasonable safety profile in smaller early studies to move into the next stage of development. It is not uncommon to see compounds selected in a smaller study show poor efficacy results in a subsequent larger study.

{ 
In a phase 2 study, anifrolumab showed an impressive improvement in systemic lupus erythematosus (SLE) responder index 4 (SRI4) at 52 weeks compared to placebo (62.6\% for anifrolumab 300 mg vs. 40.2\% for placebo; p<0.001)\cite{furie2017anifrolumab}. However, recently 2 phase 3 studies showed less impressive results. In TULIP-1 Study, the response rates were 36.2\% for anifrolumab 300 mg vs. 40.4\% for placebo \cite{furie2019type}. In TULIP-2 Study, the response rates were 55.5\% and 37.3\% for anifrolumab 300 mg and placebo treatments, respectively (p < 0.001) \cite{morand2020trial}. The first study did not reach statistical significance and neither study achieved the magnitude of the treatment effect as observed in the phase 2 study.  
 No clear differences in the study design and study population could explain the sharp contrast of results between the phase 2 study and the first 3 study.
}


The average cost of developing a new drug that gains marketing approval is estimated to be \$1 to \$2.6 billion \cite{dimasi2016innovation, wouters2020estimated}. To improve efficiency, the most critical factor is to improve the probability of success for phase 2 and 3 development \cite{paul2010improve}. Therefore, characterizing a statistical framework that can explain the above mentioned ``sharp contrast" helps enhance quantitative decision-making during the stage-wise drug development process. 

The terminology of the proof-of-concept (PoC) study we use in the article is relative: depending on the endpoints, for some therapeutic areas, a dose-response phase 2 study may be considered a PoC study; while for some other areas, a multiple ascending dose (MAD) study may serve as a PoC study. To avoid confusion, we refer to the smaller earlier study as the Small Study, and the subsequent larger study as the Large Study throughout the article.
There are many reasons the positive result observed in a Small Study may not be replicated in the later Large Study.
For example, the populations, duration, and endpoints between the two studies may be different. In oncology, the primary endpoint in a PoC study is often progression-free survival, while the primary endpoint for a confirmatory study becomes overall survival. For diabetes, glucose may be investigated in a PoC study, while Hemoglobin A1c (HbA1c) is generally the primary endpoint for a subsequent study. While these factors are important for consideration in evaluating the difference in the results between the two studies or predicting the outcome in the Large Study based on the results from the Small Study, we will explore the fundamental reason for the difference in the estimation of the treatment effects between the two studies, assuming they are similar except for the sample size.

It has been proposed that the probability of study success (PrSS) can account for the variability in the assumed true treatment difference compared to the statistical power \cite{o2005assurance, wang2013evaluating}; However, statisticians often use a normal prior distribution with the estimated mean and variance from the Small Study (Frequentist approach), or use the posterior distribution given the data in the Small Study with non-informative prior (Bayesian approach). This approach only accounts for the variability from the Small Study, not the selection bias (only moving compounds with promising results from the Small Study to the next development stage).

The problem of selection bias or regression to the mean was described as Tweedie's formula \cite{robbins1956empirical, efron2011tweedie}. Tweedie's formula to estimate the prior density assumes the variance of measurement error was constant across observations. Its application in clinical trials has not been widely realized. 
Chuang-Stein and Kirby described the phenomenon of selection bias and regression to the mean, and provided an overview of the research in discounting the early phase study results \cite{chuang2017quantitative}. One approach is to apply an empirical discount factor for the treatment effect or to raise the bar for the criterion for moving the drug to the next development stage \cite{wang2006adapting, kirby2012discounting}. Zhang evaluates the selection bias phenomenon using a Bayesian approach assuming an informative prior distribution through simulation \cite{zhang2013evaluating}. Again, the prior was used as an empirical way of ``discounting" the results from early phase studies with no clinical meaning and how to determine the prior is not stated also in their paper.  

In this article, we will describe the aforementioned problem through a theoretical framework in the drug development context, and build the connection between Frequentist and Bayesian methods. Furthermore, we will propose a Bayesian hierarchical model to estimate the distribution for a portfolio of historical compounds, which can be used as the prior for future drug development for treating the same disease with a similar mechanism. We envision that a very important benefit of this research will be the improvement of probability of late phase study success and the reduction of the overall cost of drug development.

This article is organized as follows. In Section 2, we provide the theoretical framework to explain the systematic bias, introduced by the nature of the size of the Small Study, and the fact that we only select the promising compounds moving forward. We also describe the methods to form a prior distribution used in the framework that characterizes the treatment effect for a portfolio of similar compounds. In Section 3, we provide 2 data examples from drug development in diabetes and rheumatoid arthritis (RA) disease respectively, using the methods described in Section 2. Finally, Section 4 provides a summary and discussion of the topic.

\section{Methods}
\label{section_stat_formulation}
Assume we are interested in estimating the treatment effect comparing a new treatment (at a certain dose) to a comparator (typically placebo).
The treatment effect can be the treatment difference in means or proportions, logarithm of odds ratio, or logarithm of hazard ratio, depending on the type of variables for the outcome and study objectives. Given a compound, let $\theta$ denote the true treatment effect.
Without loss of generality, we assume a smaller value of $\theta$ means better efficacy. Consider a typical sequential clinical development program for a compound in which an exploratory Small Study is conducted first, and the subsequent Large Study is conducted upon promising results from the Small Study. Assume 
the estimators of the treatment difference for the Small Study and Large Study are distributed from
\begin{equation} \label{eq:S}
\hat{\theta}_S | \theta \sim \mathcal{F}_S(\theta)
\end{equation}
and
\begin{equation} \label{eq:L}
\hat{\theta}_L | \theta \sim \mathcal{F}_L(\theta),
\end{equation}
respectively. We further assume the compound being studied is from a portfolio of candidate treatments with the true treatment effect distributed from 
\begin{equation} \label{eq:theta}
\theta \sim \mathcal{F}_\theta.
\end{equation}
We assume 
\begin{enumerate}
  \item $\hat{\theta}_S \perp \hat{\theta}_L | \theta$, i.e., the conditional independence between $\hat{\theta}_S$ and $\hat{\theta}_L$ given the treatment effect $\theta$ for the compound under investigation.
  \item $E[\hat{\theta}_S|\theta] = E[\hat{\theta}_L|\theta] = \theta$. Although sometimes the estimators are only asymptotically unbiased given $\theta$, we assume the conditional unbiasedness for the convenience of calculation and argumentation. 
\end{enumerate}  

\subsection{The probability of observing a large treatment effect in a small PoC study}
We now illustrate the impact of the variability in $\hat{\theta}_S$ on the probability of meeting a desired threshold value for the treatment difference through a normal prior distribution:
\[
\hat{\theta}_S | \theta \sim N(\theta, \sigma_s^2).
\]

\begin{figure}[h!tb] \centering
\rotatebox{0}{\resizebox*{4in}{4in}{\includegraphics*{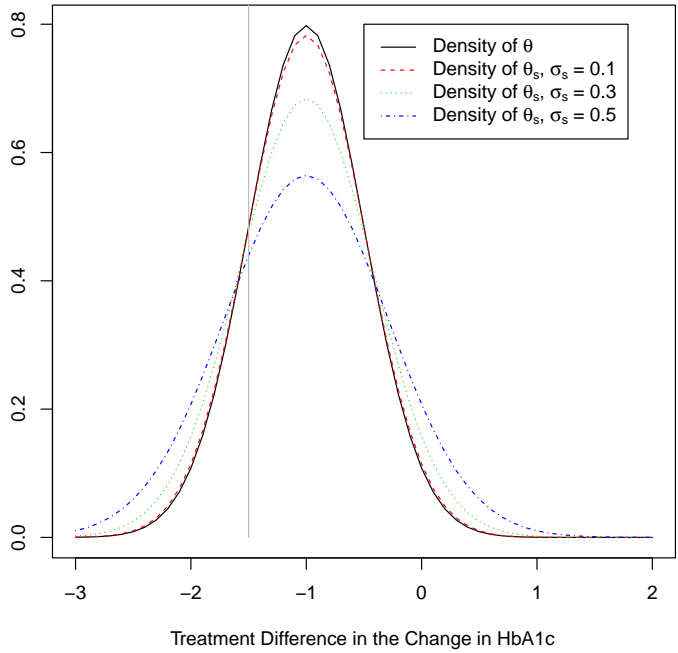}}}
\caption{The distribution (density) of the true treatment effect ($\theta$) and the estimator for the treatment effect in the Small Study $\hat {\theta}_S$.}
\label{plot:me}
\end{figure}

For example, suppose we intend to develop an anti-diabetes treatment and the primary endpoint in the Small Study is the change in HbA1c from baseline at week 12. The desired treatment effect for the new treatment versus placebo (difference in group means) is at least as good as $\delta = -1.5\%$. Figure \ref{plot:me} shows the relationship between the probability of meeting the threshold value [$\Pr(\hat{\theta}_S
\le \delta$)] and the standard deviation, $\sigma_s$, for $\theta \sim N(-1.0, 0.5)$. We can see that $\Pr(\hat{\theta}_S
\le \delta$) increases as $\sigma_s$ goes up. This means the smaller the sample size, the more likely we would be to observe a promising treatment effect for this particular compound in the portfolio.

\subsection{Define and adjust for the selection bias from a frequentist approach} \label{sec:method_frequentist}
Although $\hat{\theta}_S \perp \hat{\theta}_L | \theta$, an underlying distribution for $\theta$ imposes an unconditional correlation between $\hat{\theta}_S$ and $\hat{\theta}_L$. A joint distribution between $\hat{\theta}_S$ and $\hat{\theta}_L$ is shown below:
\begin{align}
f(\hat{\theta}_S, \hat{\theta}_L) & = \int f(\hat{\theta}_S, \hat{\theta}_L, \theta) d\theta \nonumber \\
                                  & = \int f(\hat{\theta}_S, \hat{\theta}_L | \theta)f(\theta)d\theta \nonumber \\
				  & = \int f(\hat{\theta}_S | \theta) f(\hat{\theta}_L | \theta)f(\theta)d\theta.
\end{align}
The conditional distribution $f(\hat{\theta}_L | \hat{\theta}_S)$ can subsequently be derived accordingly:
\begin{equation} \label{eq:L_on_S}
f(\hat{\theta}_L | \hat{\theta}_S) = \frac{f(\hat{\theta}_S, \hat{\theta}_L)}{f(\hat{\theta}_S)}.
\end{equation}
The drug development is a stage-wise process: we only move the compound to the Large Study if the Small Study shows a reasonably promising result (e.g., $\hat \theta_{S}<\delta$). Therefore, even though $\hat \theta_{L}$ itself is an unbiased estimator for $\eta$, $\hat \theta_{L}$ given $\hat \theta_{S}<\delta$ is no longer an unbiased estimator for $\eta$.

We now evaluate two quantities of interest from the Small Study based on the conditional distribution:
\begin{enumerate}
  \item The mean treatment difference for the Large Study conditional on the result from the Small Study, i.e., $E(\hat{\theta}_L| \hat{\theta}_S)$.
     Since $E(\hat{\theta}_L| \hat{\theta}_S)$ is a function of $\hat{\theta}_S$, it has no selection bias. Therefore, $E(\hat{\theta}_L| \hat{\theta}_S)$ is an unbiased estimator for $\eta$, which can be seen by $E[E(\hat{\theta}_L| \hat{\theta}_S)]=E(\hat{\theta}_L) = \eta$.
  \item The conditional probability of achieving the desired treatment effect for the Large Study conditional on the result from the Small Study, i.e., $\Pr(\hat{\theta}_L < \delta |  \hat{\theta}_S)$. It can also been seen that $E\{\Pr(\hat{\theta}_L < \delta |  \hat{\theta}_S)\} = \Pr(\hat{\theta}_L < \delta)$.
\end{enumerate}

The {\em bias} (sometimes also called {\em discount factor}) for using $\hat{\theta}_S$ to estimate the expected treatment effect for the Large Study given the Small study is defined by

\begin{equation} \label{eq:bias}
\Delta(\hat{\theta}_S) = E(\hat{\theta}_L | \hat{\theta}_S) - \hat{\theta}_S.
\end{equation}
The bias is computed based on the information from the Small Study and the information from the portfolio, but independent of the estimate from the Large Study. It represents the amount of adjustment or discount factor we shall apply to the treatment effect estimate from the Small Study in order to have a more realistic view of the expected treatment effect from the Large Study.

\begin{figure}[h!tb] \centering
\rotatebox{0}{\resizebox*{4in}{4in}{\includegraphics*{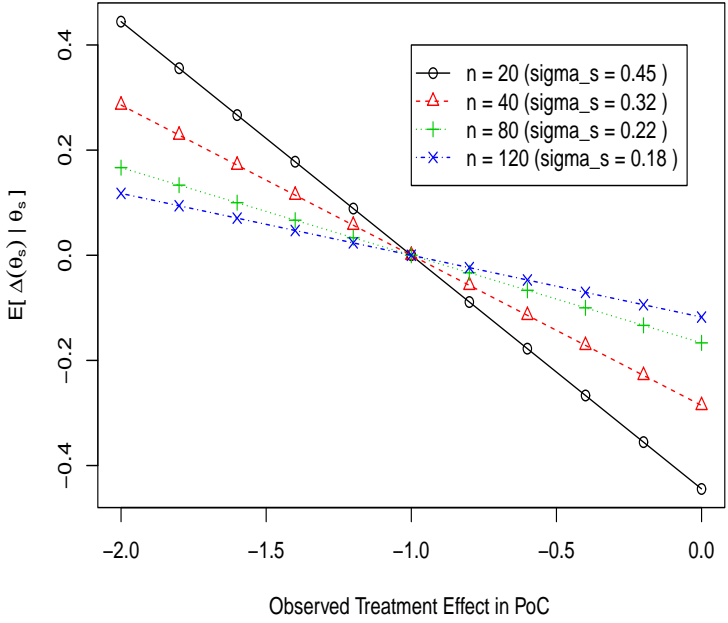}}}
\caption{ \small
The relationship between observed treatment effect in the Small Study (x-axis) and the difference in the conditional treatment effect between the Large Study and Small Study (y-axis), assuming $\eta=-1.0$ and $\sigma=1.0$.
}
\label{plot:bias}
\end{figure}

To have a more concrete perspective on the definition of ``bias", let us assume a special case where $\mathcal{F}_S, \mathcal{F}_L$, and $\mathcal{F}_\theta$ are normal distributions such that
\begin{equation} \label{eq:Sn}
\hat{\theta}_S|\theta \sim N(\theta, \sigma_S^2),
\end{equation}
\begin{equation} \label{eq:Ln}
\hat{\theta}_L|\theta \sim N(\theta, \sigma_L^2),
\end{equation}
and
\begin{equation} \label{eq:mu}
\theta \sim N(\eta, \sigma^2),
\end{equation}
{ where $\eta$ and $\sigma^2$ are the mean and variance for the prior distribution of $\theta$.} 
The unconditional distribution of $(\hat{\theta}_S,\hat{\theta}_L)$ is easily derived as
\begin{equation} \label{eq:SL}
\left[ \begin{array}{c} \hat{\theta}_S \\ \hat{\theta}_L \end{array} \right] \sim N\left(
\left[\begin{array}{c} \eta \\ \eta \end{array} \right],
\left[\begin{array}{cc} \sigma_S^2+\sigma^2 & \sigma^2\\ \sigma^2 & \sigma_L^2+\sigma^2 \end{array} \right]
\right).
\end{equation}
The conditional distribution of $\hat{\theta}_L$ given $\hat{\theta}_S$ is
\begin{equation} \label{eq:conditional}
\hat{\theta}_L | \hat{\theta}_S = s \sim N(\eta+(\sigma_S^2+\sigma^2)^{-1}\sigma^2(s-\eta), (\sigma_L^2+\sigma^2) - (\sigma_S^2+\sigma^2)^{-1}\sigma^4).
\end{equation}
According to the definition by (\ref{eq:bias}), the bias is given by
\begin{equation} \label{eq:bias1}
\Delta(s) = \eta+(\sigma_S^2+\sigma^2)^{-1}\sigma^2(s-\eta) - s = (\sigma_S^2+\sigma^2)^{-1}\sigma_S^2(\eta-s).
\end{equation}

Figure \ref{plot:bias} shows the bias related to the observed treatment effect in the Small Study for various sample sizes of the Small Study.
 {In this plot, we mimic the variable of the change in HbA1c in anti-diabetes drug development, assuming the mean and standard deviation for the prior distribution are -1\% and 1\%, respectively. In anti-diabetes treatment PoC or phase 2 studies, the sample size is generally between 20 and 100 per treatment arm.}
Since $\eta$ is the mean of a portfolio of candidate drugs of the same class and the sponsors tend to pick up the compounds with promising results from the Small Study to move forward, $s$ is generally smaller than $\eta$ (keep in mind smaller values mean better efficacy). Then, the bias is always positive, and the treatment effect observed in the Large Study is very likely worse than that in the Small Study. The more promising and more variable the Small Study result is, the larger the difference between the expected treatment effects from the 2 studies will be. Therefore, more discount shall be applied towards the observed treatment effect in the Small Study in the planning of the next study, to offset the magnitude and variability in estimating $\hat{\theta}_S$.

\begin{figure}[h!tb] \centering
\rotatebox{0}{\resizebox*{4in}{4in}{\includegraphics*{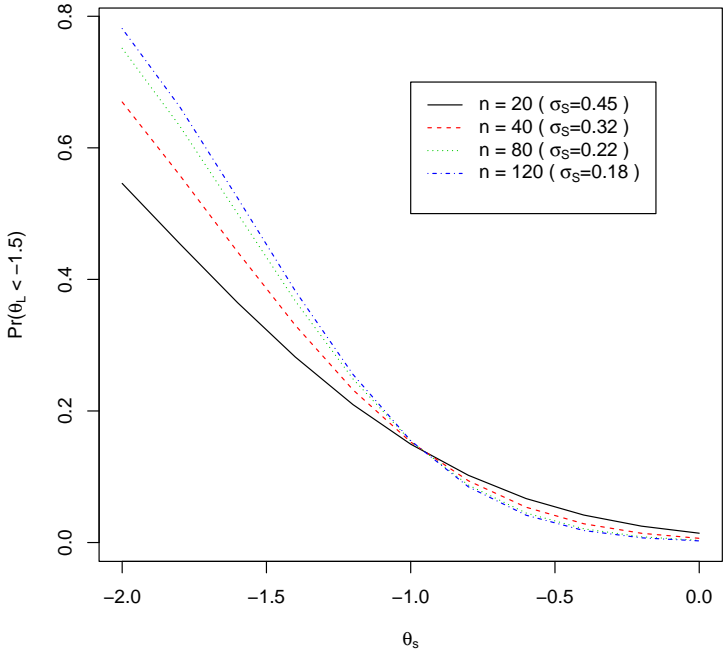}}}
\caption{ \small
The relationship between the observed treatment difference estimate from the Small Study and the conditional probability of the success of the subseqeunt Large Study  (assuming $\eta=-1.0$, $\sigma=1.0$, and the sample size is 400 for the Large Study).
}
\label{plot:cp}
\end{figure}

The conditional distribution of $\hat{\theta}_L$ given $\hat{\theta}_S$ also enables us to estimate the probability of observing the desired efficacy conditional on the result from the Small Study. Figure \ref{plot:cp} depicts the relationship between the observed treatment effect from the Small Study ($\hat{\theta}_S$) and the conditional probability of the success of the Large Study for different standard error of the observed treatment difference in the Small Study (with a similar setting as Figure \ref{plot:bias}).
Again, the more promising the observed treatment effect is in the Small Study, the more likely the next study is to succeed. However, given any fixed value of $\hat{\theta}_S$, as the standard error of the estimate increases from 0.18 to 0.45, the conditional probability of the success of the Large Study falls by $20$-$30$\%.

From (\ref{eq:conditional}), the adjusted estimator of the expected treatment effect for the Large Study is readily available:
\[
E[\hat{\theta}_L | \hat{\theta}_S = s] = \eta+(\sigma_S^2+\sigma^2)^{-1}\sigma^2(s-\eta).
\]
Although it is called a special case when $\mathcal{F}_S, \mathcal{F}_L$, and $\mathcal{F}_\theta$ are normal distributions, this is, in fact, quite general and can apply to a wide variety of scenarios in clinical development. By central limit theorem, with moderate or large sample size, both $\hat{\theta}_L$ and $\hat{\theta}_S$ approximately follow normal distributions regardless of the distribution of their associated outcomes. This is not very different from the estimation and inferences in the Frequentist approach, where the normal approximation is commonly used. The only distribution that could be very non-Gaussian is the prior distribution $\mathcal{F}_\theta$, which depends on the nature of the portfolio of drugs under investigation.

\subsection{Understand and adjust for the bias from the perspective of Bayesian statistics}
As Efron \cite{efron2011tweedie} pointed out, Bayesian method inherently prevents the bias if the appropriate prior distribution is used. Let $f(\theta|\hat \theta_S)$ be the posterior distribution given the estimator $\hat \theta_S$ with a prior distribution of (\ref{eq:theta}). The conditional expectation $E(\hat{\theta}_L| \hat{\theta}_S)$ is equivalent to the Bayesian posterior mean for the Small Study with the prior distribution of (\ref{eq:theta}). This can be seen by
\begin{align}
E(\hat{\theta}_L | \hat{\theta}_S) = E\{E(\hat{\theta}_L  |  \hat{\theta}_S, \theta)|\hat{\theta}_S\}
= E\{E(\hat{\theta}_L | \theta)|\hat{\theta}_S\}
= E(\theta |\hat{\theta}_S).
\end{align}
Similarly, 
\begin{align}
\Pr(\hat{\theta}_L < \delta |  \hat{\theta}_S) = E\{\Pr(\hat{\theta}_L < \delta |  \hat{\theta}_S, \theta)|\hat{\theta}_S\}
= E\{\Pr(\hat{\theta}_L < \delta | \theta)|\hat{\theta}_S\}
\end{align}
is the posterior expectation of $E\{\Pr(\hat{\theta}_L < \delta | \theta)\}$ given the estimator from the Small Study. 

Often the sponsor who conducts the Small Study has the individual data $Y_S$. Then, a fully Bayesian method based on the observed data (instead of the estimator $\hat \theta_S$) can be used to estimate the posterior distribution $f(\theta|Y_S)$. The adjusted point estimator is 
\begin{align}
E(\hat{\theta}_L | Y_S) = E(\theta |Y_S),
\end{align}
and the adjusted probability of meeting a threshold for the treatment effect is
\begin{align}
\Pr(\hat{\theta}_L < \delta |  Y_S) = E\{\Pr(\hat{\theta}_L < \delta | \theta)|Y_S\}.
\end{align}

When the individual patient data $Y_S$ are available, the estimation based on the posterior distribution $f(\theta|Y_S)$ is more natural, and may be more accurate compared to $f(\theta|\hat \theta_S)$ when the distribution of $\hat \theta_S$ is not exactly normally distributed; while the estimation based on the posterior distribution given $\hat \theta_S$ provides an advantage when only the estimate for the Small Study is available. 

\subsection{Modeling the distribution of $\theta$}
\label{subsec:theta}
The estimation of the distribution of $\theta$ is the key to the understanding of the portfolio performance in both Frequentist and Bayesian methods. In this section, we will use a Bayesian hierarchical model to form the prior distribution.

Following up the special case in Section~\ref{sec:method_frequentist}, the estimates of $\eta$ and $\sigma^2$ are required to carry out the bias adjustment as defined by (\ref{eq:bias1}). We propose to construct a hierarchical model for the estimation of parameters $\eta$ and $\sigma^2$. At any given point in time, we use the study data on the compound portfolio accumulated so far to infer the parameter of the distribution of $\theta$. As the portfolio is expanding with time, we should update the inference when more relevant study information becomes available.

Suppose we have information for $I$ compounds. For the $i^{th} (1 \le i \le I)$ compound, there are $m_i$ studies. Note that it is important to include data 
from both positive and negative studies, and both early and late phase studies. For some studies, the population, endpoint or study duration may be different from what we desire, either these studies are excluded or the treatment effect for the desired population, endpoint and study duration is estimated with additional modeling and extrapolations. The estimates from the studies can be modeled as
\begin{equation} \label{eq:pS}
\hat{\theta}_{ij}|\theta_i \sim N(\theta_i, {\sigma}_{ij}^2), \;\; i=1,2,\ldots,I; \;\; j=1,2,\ldots,m_i
\end{equation}
and
\begin{equation} \label{eq:pmu}
\theta_i \sim N(\eta, \sigma^2).
\end{equation}

By assigning prior distributions for $\eta$ and $\sigma^2$, the inference on $\eta$ and $\sigma^2$ can be easily obtained from the maximum likelihood estimation or using Bayesian hierarchical modeling framework. Note Tweedie's formula cannot be used here as the variances $\sigma_{ij}^2$ are not constant \cite{robbins1956empirical, efron2011tweedie}.  
We suggest a non-informative or weakly informative prior for the distribution of $\eta$ and $\sigma^2$, where for example, $\eta \sim N(0, 1000);\sigma^2 \sim \text{Inverse Gamma } (0.001, 0.001)$. Standard softwares are readily available to draw the posterior inference based on the defined hierarchical modeling, such as JAGS, WinBUGS and STAN. Once the posterior estimates for $\eta$ and $\sigma^2$ are obtained, they can be fed into (\ref{eq:bias1}) to gauge the bias for any compound in the portfolio finishing the Small Study with a promising result, and construct an adjusted estimator for the assessment of efficacy of this compound and planning for the next study. 

{ 
The method of estimating $\mathcal{F}_\theta$ can be easily generalized to other models or assuming a non-Gaussian distribution. For example, a more complex approach to model the prior is to include the within-compound between-study variability. Specifically, one can replace equation (\ref{eq:pS}) with
\begin{equation} \label{eq:pS_ij}
\hat{\theta}_{ij}|\theta_{ij} \sim N(\theta_{ij}, {\sigma}_{ij}^2), \; \theta_{ij}|\theta_i \sim N(\theta_{i}, {\sigma}_{i}^2),
\;\; i=1,2,\ldots,I; \;\; j=1,2,\ldots,m_i,
\end{equation}
where $\theta_{ij}$ the mean for study $j$ and compound $i$, ${\sigma}_{ij}^2$ is the variance of $\hat \theta_{ij}$ given $\theta_{ij}$,  
$\theta_{i}$ is the mean for compound $i$, and ${\sigma}_{i}^2$ is the between-study variance for compound $i$. 
In some cases it may be difficult to estimate the within-compound between-study variability when there is only 1 Small Study for a compound. 
}
\section{Real Data Examples}
In this section, we illustrate the application of bias adjustment through real data examples in two therapeutic areas: diabetes and immunology. In the diabetes therapeutic area, we considered the endpoint of body weight for the class of incretins as anti-diabetic drugs for patients with Type-2 diabetes while in the immunology therapeutic area, we focused on the treatment against the disease of rheumatoid arthritis (RA).

\subsection{Diabetes Therapeutic Area}
Due to confidentiality requirements, the compounds have been de-identified. The proposed method in Section~\ref{subsec:theta} is implemented to model the distribution of the treatment effect $\theta$ in the portfolio of candidate compounds. To that purpose, we gathered and used the available study data (from publication or conference presentations) with regards to the effect of incretins on weight loss across pharmaceutical companies. In this real data application, the Small Study refers to the phase 1b MAD study while the Large Study corresponds to the phase 2 study. Again, we assume normal distributions for $\mathcal{F}_S, \mathcal{F}_L$, and $\mathcal{F}_\theta$. {  In this illustrative example, the outcome is the change in body weight from baseline to 4 weeks}. There are 5 compounds for which we know the estimates and their standard errors for both phase 1b and phase 2 studies, so that we can compare the bias-adjusted estimate and the observed treatment effect in phase 2 study. {  In all studies for all 5 compounds, the population was similar: patients with Type-2 diabetes with inadequate glycemic control by diet and exercise or the treatment of metformin}.
The application follows the below procedure:

\begin{enumerate}
  \item Use the method in Section~\ref{subsec:theta} to model the distribution $\mathcal{F}_\theta$ with all available information.
  \item Apply the bias adjustment as in (\ref{eq:bias1}) to those 5 compounds, compute the new estimate $E[\hat{\theta}_L | \hat{\theta}_S = s]$, and compare with the observed treatment effect for the phase 2 study.
\end{enumerate}


\begin{figure}[h!tb] \centering
\rotatebox{0}{\resizebox*{4.8in}{4in}{\includegraphics*{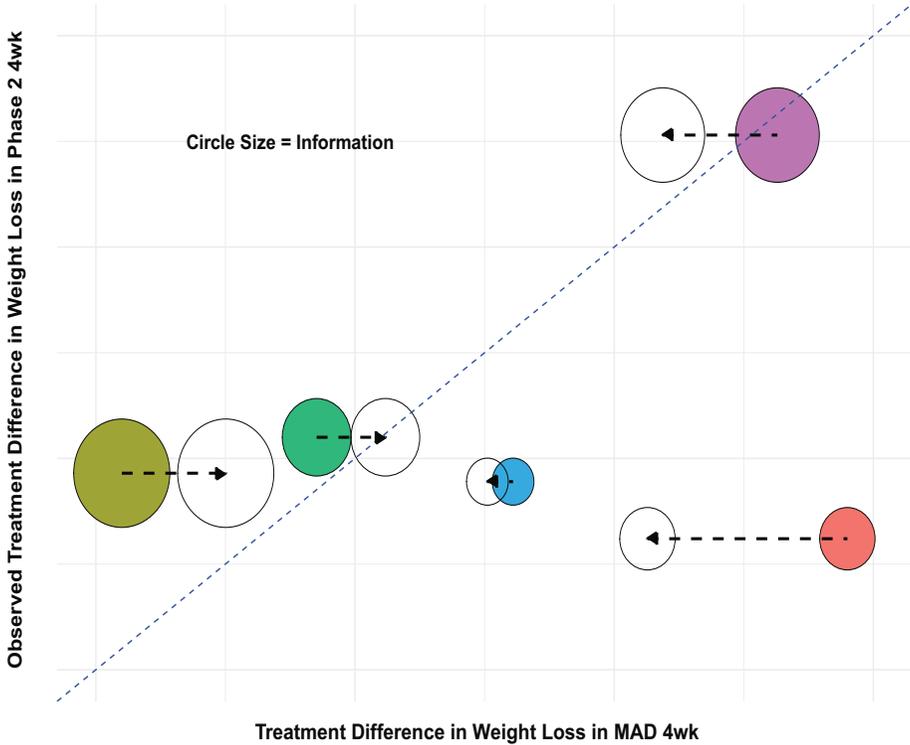}}}
\caption{ \small
The application of the bias adjustment in estimating the treatment effect on the change in body weight from baseline to 4 weeks for anti-diabetic incretin drugs. The x-axis for the solid circle indicates observed (unadjusted) treatment effect in Small Study and the x-axis for the hollow circle indicates adjusted treatment effect in Small Study.
}
\label{plot:diabetes}
\end{figure}

The actual availability of the Large Study (phase 2 study) results allow for a comparison of the un-adjusted and adjusted estimates ($\hat{\theta}_{S}$ and $E[\hat{\theta}_L | \hat{\theta}_S = s]$, respectively) with the observed estimates of treatment effect in phase 2 study. Figure~\ref{plot:diabetes} shows a graphical display of the unadjusted treatment effect based on the Small Study ($\hat{\theta}_{S}$) versus the observed treatment effect in the Large Study (solid circles), and the adjusted treatment effect based on the Small Study ($E[\hat{\theta}_L | \hat{\theta}_S = s]$) versus the observed treatment effect in the Large Study (hollow circle) for all 5 incretin compounds. The circle size (area) is proportionate to the information of the estimates (i.e., one over the square of the standard error). {  Most data used in this figure are confidential and have not been published, so we removed the scales for the x-axis and y-axis. For the illustration of effectiveness of the bias adjustment, the relative, but not the absolute scale, is sufficient.}
Ideally, we expect the solid circles (unadjusted estimates) to be distributed symmetrically across the 45-degree line; however, most of the time the sponsors are only willing to move the compounds to the next developmental stage when they exhibit promising results in the Small Study. This results in an uneven distribution of the solid circles with more circles in the lower right quadrant and fewer circles in the upper left area, a phenomenon of selection bias. 

The adjusted estimates are closer to the observed treatment effect in phase 2 studies, as indicated by all the hollow circles with the exception of
the purple compound moving away from the 45-degree line. This is not surprising because both the adjusted estimates from phase 1b studies (also called Bayesian shrinkage estimates) and the unadjusted estimate from phase 2b studies were subject to variability and, more importantly the Bayesian shrinkage estimator provides better overall estimation (e.g., in terms of mean squared errors) but may be biased if conditional on individual compounds. The estimate for the orange compound had relatively large variance for the treatment effect estimate (as indicated by a small circle) and a large treatment effect (much more than $\eta$). Therefore, the circle for the orange compound had the most shift to the left. On the other hand, there was also one compound (the leftmost) with almost no weight loss based on the phase 1b study, while some weight loss was observed in a subsequent Large Study. This is rare in practice but it is possible for several reasons, for example, the primary outcome upon which the decision is made may not be weight loss. {  In addition, Figure~\ref{plot:diabetes} shows the adjustment may not always make the treatment effect smaller. For the 2 leftmost compounds, the adjusted treatment effect was larger compared to the unadjusted one. This phenomenon is consistent with the property of the Bayesian shrinkage estimator, which shrinks the estimates to the center of the prior distribution. }

\subsection{Immunology Therapeutic Area}
In this section, we further illustrate how the bias-adjusting method could be implemented in an RA example.
While there are a handful of compounds approved for RA, it is not uncommon that a compound which is promising in a phase 2 study fails in phase 3. 
In this example, the Small Study refers to the phase 2 while the Large Study corresponds to the phase 3 study. The outcome is set to be ACR20 (whether a patient has $\geq20\%$ improvement in ACR [American College of Rheumatology] assessment) at week 12, and the treatment effect $\theta$ is defined as the difference in ACR20 response rates between an experimental treatment and placebo arms. 
As treatment effect varies significantly across different subpopulations, we only focus on studies with populations that have had an inadequate response to methotrexate (MTX). As previously mentioned, normal distributions for $\mathcal{F}_S, \mathcal{F}_L$, and $\mathcal{F}_\theta$ are assumed throughout this section.

To estimate the portfolio distribution of $\theta\sim N(\eta, \sigma^2)$, we perform a systematic review of literature and select RA clinical trials based on two criteria: (i) double-blind, placebo-controlled RA trials with ACR20 results reported at week 12; (ii) $>50\%$ of enrolled patients have inadequate-response to MTX. Ultimately, 48 phase 2 trials are selected. We apply the method in Section~\ref{subsec:theta} using published estimates and standard errors in the selected phase 2 trials. 
The prior distribution is therefore estimated such that $\theta \sim N(0.244,0.14^2)$, which is then used to calculate the adjusted treatment effect $E[\hat{\theta}_L | \hat{\theta}_S = s]$. Fourteen out of 48 phase 2 trials have been corresponded to 15 phase 3 trials, all of which share similar population and background therapies as those in selected phase 2 trials. 
Note that a phase 2 trial may be matched with more than one phase 3 trials. In this case, we used the meta-analysis to pool the results from multiple phase 3 trials and treated as one large phase 3 study. 
{  Overall, there were 9 compounds and 24  compound-dose levels (``treatments") included in the analysis. }

Figure 5 
represents the unadjusted treatment effect $\hat{\theta}_{S}$ based on the phase 2 study versus the observed treatment effect in the phase 3 study (solid circles), and the adjusted treatment effect based on the phase 2 study $E[\hat{\theta}_L | \hat{\theta}_S = s]$ versus the observed treatment effect in the phase 3 study (hollow circle). {  Each circle represents 1 compound-dose level.} The circle size (area) is proportionate to the information of the estimates (i.e., one over the square of the standard error).
The observed treatment effects $\hat{\theta}_{S}$ in phase 2 range from $0.1$ to $0.75$, while phase 3 results are more stable and treatment effects $\hat{\theta}_{L}$ are within the range $0.2$ to $0.4$. Again, most of the solid circles fall under the 45-degree line, which means that in general phase 3 results appear worse than originally reported phase 2 results. This is another example of selection bias.
Compared to the observed phase 2 treatment difference estimator $\hat{\theta}_S$, the bias-adjusted estimator $E[\hat{\theta}_L | \hat{\theta}_S = s]$ is closer to its observed phase 3 results $\hat{\theta}_L$.
The plot also shows the magnitude of bias-adjustment through the length of the arrow between each pair of solid and hollow circles. The longer the arrow, the more bias-adjustment is present. Such bias-adjustment is associated with two terms:  (i) difference between $\eta$ and the observed phase 2 result $\hat{\theta}_S$: the closer the observed phase 2 results are to $0.244$, the smaller adjustment shall be applied; (ii) phase 2 sample size: larger phase 2 sample size leads to a smaller adjustment, as indicated in (13).



\begin{figure} \label{plot:RA}
    \centering
    \includegraphics[scale=0.7]{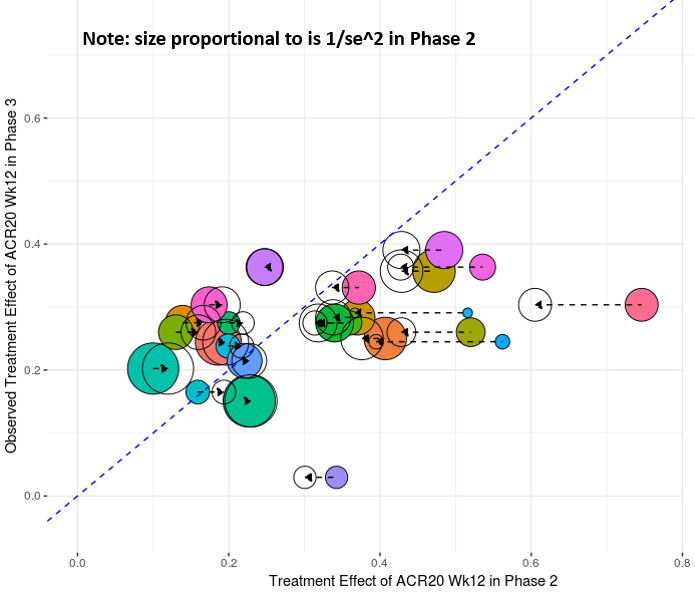}\\
    \caption{ \small
The application of the bias adjustment in estimating the treatment effect of selected RA compounds. The x-axis for the solid circle indicates observed (unadjusted) treatment effect in Small Study and the x-axis for the hollow circle indicates adjusted treatment effect in Small Study.
}
\end{figure}

\section{Summary and Discussion}
\label{sec:discussion}
Selection bias or regression to the mean phenomenon has been observed in the past, and some research has been done in this area. In this article, we made new contributions in four aspects: the clinical meaning of the prior, the connection between the estimation bias due to selection bias and the Bayesian posterior mean, the use of hierarchical modeling to estimate the distribution of the underlying treatment effect for a portfolio of compounds, which can be used as the prior distribution to adjust the treatment effect for current and future studies, and the role of estimation variability in the Small Study in the estimation bias.

Although prior distribution has been used in the estimation of posterior mean to account for selection bias, the clinical meaning of prior distribution was not clearly defined \cite{zhang2013evaluating}. In this article, we defined that the prior represents the distribution of the underlying treatment effect of a portfolio of similar compounds. The clarification of the clinical meaning of prior distribution allows us to develop a Bayesian hierarchical model to estimate the prior distribution. We start from the commonly understood bias as $E[\hat \theta_L|\hat \theta_S] - \hat \theta_S$ and showed this is equivalent to $E[\theta|\hat \theta_S] - \hat \theta_S$, the ``regression to the mean" (discount) effect from the Bayeisan framework, if $\theta$ is the treatment effect. While Chuang-Stein and Kirty consider the discount based on the Bayesian framework as a different definition of ``regression to the mean" \cite{chuang2017quantitative}, we established the equivalence of discount between Frequentist and Bayesian methods. We also further quantified the variability in the estimator for the Small Study on the probability of achieving a large promising treatment effect, as well as on the bias. The larger the estimation variability and the larger the effect size in the Small Study, the more discount should be applied to adjust for the bias in the estimator for the Large Study. Therefore, special caution should be taken when a very promising signal is observed from a very small study.

This phenomenon may be perplexing. For a single asset, $\hat \theta_S$, the estimator from the Small Study, is unbiased for treatment effect since $E(\hat \theta_S|\theta) = \theta$. However, from a portfolio perspective, the distribution of the observed treatment effect $\hat \theta_S$ is more variable than the distribution of true $\theta$, but $\hat \theta_S$ is still an unbiased estimator for the mean of true treatment effect for the portfolio. This can be seen from the theoretical perspective:
\[
Var(\hat \theta_S) = E[Var(\hat \theta_S|\theta)] + Var[E(\hat \theta_S|\theta)] = E[Var(\hat \theta_S|\theta)] + Var(\theta) > Var(\theta)
\]
and
\[
E(\hat \theta_S) = E[E(\hat \theta_S|\theta)] = E[\theta].
\]
A Bayesian shrinkage estimator will provide a more realistic estimation of the distribution of $\theta$.
When only promising compounds (e.g., $\hat \theta_S < \delta$) based on the Small Study are moved forward for the next stage of development, a selection bias is introduced. In the presence of such selection bias, $\hat \theta_S$ is also biased for $\theta$. This can be seen by
\[
E(\hat \theta_S | \hat \theta_S < \delta) \ne E(\theta).
\]

In this article, we provided two examples with the treatment effects being the difference in means and difference in proportions and a normal prior. The theoretical framework in this article can be applied to other contrasts of treatment effect (e.g., logarithm of the odds ratio or logarithm of the hazard ratio), since all these estimators approximately follow normal distribution. The prior distribution for $\theta$ does not have to be a normal distribution. For example, the prior distribution may be a mixture of two normal distributions representing two sets of compounds with little and reasonable treatment effect respectively, or contain a mass on the treatment effect of zero to account for drugs with no treatment effect.
{ 
Selection of studies/compounds to form the prior distribution $\mathcal{F}_\theta$ is important in application of the proposed method to adjust for the selection bias. For a new compound that is not novel (i.e., data are available for similar compounds), the data from compounds in the same or similar class should be used to form the prior. For a compound with a novel target (i.e., no data are available for similar compounds), we may assume this compound comes from a distribution of existing classes of treatments. For such compounds, a prior distribution to characterize different classes of treatments for the same disease can be used: either a prior constructed by treating each class as one data point (mean treatment effect for each class based on meta-analysis) as described in Equation (\ref{eq:pS}) or a hierarchical model considering the between-study, compound and class variabilities [e.g., Equation (\ref{eq:pS_ij})].
}

{ 
In addition to sample size, there may be many differences between Small and Large Studies, including but not limited to population, endpoint, duration, and dose. In this article, we proposed a method to predict mean treatment effect of Large Study based on Small Study in consideration of adjusting for the selection bias only, assuming Small and Large Studies are otherwise similar except for sample size. Although we generally try to make phase 2 studies more translatable to phase 3 studies, considerable difference may still exist between phase 2 and 3 studies, possibly due to the safety consideration, limited knowledge on the candidate treatment, and financial consideration. In this case, one should take a 2-step approach for predicting the treatment effect of the candidate treatment in phase 3 studies. First, a model to account for the differences between phase 2 and 3 studies should be built to predict phase 3 results without adjusting for the selection bias. Then, the method proposed in this article can be applied to the predicted outcome in Step 1 to adjust for the selection bias. The prediction model as well as the prior for adjusting for selection bias should be pre-specified and developed before the completion of Small Study so that (1) the prediction model and the prior can be agreed upon before knowing the Small study results to avoid bias due to subjectivity, and (2) the adjusted treatment effect can be estimated expeditiously right after the Small Study results become available.
}

\section*{acknowledgements}
We thank Dr. Ilya Lipkovich and Dr. Michael Sonksen for his scientific review of this manuscript and useful comments.

\section*{conflict of interest}
All authors are employees and stock holders of Eli Lilly and Company.

\printendnotes

\bibliography{citation}



\end{document}